\documentclass[preprint, 3p, times, twocolumn]{elsarticle}
\usepackage{physics, siunitx, lineno, upgreek, amsmath, graphicx, epsfig, booktabs, threeparttable, textcomp}
\usepackage{hyperref}
\modulolinenumbers[5]
\journal{Journal of Magnetic Resonance}

\begin{document}

\begin{frontmatter}

\title{Single spin magnetic resonance}

\author[PI3,MPI]{J\"{o}rg Wrachtrup}
\ead{wrachtrup@physik.uni-stuttgart.de}
\author[PI3]{Amit Finkler}
\ead{a.finkler@physik.uni-stuttgart.de}

\address[PI3]{3.\ Physikalisches Institut, Universit\"{a}t Stuttgart, Pfaffenwaldring 57, 70569 Stuttgart, Germany}
\address[MPI]{Max Planck Institute for Solid State Research, Heisenbergstr.\ 1, 70569 Stuttgart, Germany}

\begin{abstract}
Different approaches have improved the sensitivity of either electron or nuclear magnetic resonance to the single spin level. For optical detection it has essentially become routine to observe a single electron spin or nuclear spin. Typically, the systems in use are carefully designed to allow for single spin detection and manipulation, and of those systems, diamond spin defects rank very high, being so robust that they can be addressed, read out and coherently controlled even under ambient conditions and in a versatile set of nanostructures. This renders them as a new type of sensor, which has been shown to detect single electron and nuclear spins among other quantities like force, pressure and temperature. Adapting pulse sequences from classic NMR and EPR, and combined with high resolution optical microscopy, proximity to the target sample and nanoscale size, the diamond sensors have the potential to constitute a new class of magnetic resonance detectors with single spin sensitivity. As diamond sensors can be operated under ambient conditions, they offer potential application across a multitude of disciplines. Here we review the different existing techniques for magnetic resonance, with a focus on diamond defect spin sensors, showing their potential as versatile sensors for ultra-sensitive magnetic resonance with nanoscale spatial resolution.\\

\noindent \textcopyright\ 2016 This manuscript version is made available under the \href{http://creativecommons.org/licenses/by-nc-nd/4.0/}{CC-BY-NC-ND 4.0 license}.
\end{abstract}

\begin{keyword}
nuclear magnetic resonance, electron spin resonance, nitrogen-vacancy center in diamond
\end{keyword}

\end{frontmatter}

%\linenumbers

\section{Introduction}\label{intro}
Magnetic resonance, either as electron paramagnetic resonance (EPR) or as nuclear magnetic resonance (NMR), is one of the most abundant analytical and imaging techniques. Because it is sensitive to transitions between nuclear states, the attainable spectral resolution is without peer. EPR spectra often show resolution of hyperfine coupling to nuclei giving access to information on chemical composition, electron density and geometrical structure. NMR, on the other hand, shows unprecedented chemical specificity through analyzing chemical shift and $J$-coupling data. Nevertheless, both techniques are limited in sensitivity since they mostly rely on inductive interaction of spins with the detection device being a pickup coil in the case of NMR setups - which typically is challenged by thermal noise in the detection signal. More specifically, inductive detection is well suited to measuring large magnetic moments rather than single spins from basic geometric considerations. The inductive voltage signal through a surface is proportional to the time derivative of the magnetic flux flowing through it. The latter, in turn, is proportional to the number of magnetic flux lines that pass through a specific surface. If one considers a sample inside a pick-up coil then the inductive signal will only be generated by those field lines closing outside the loop. Given a certain loop size, the magnitude of the signal will decrease with shrinking sample size. For a single spin the inductive signal will thus be drastically reduced unless the loop size is reduced at the same time. Since loop sizes below a few tens of \si{\micro\second} are limited by self-induction, no single spin detection is feasible with this technique. The state-of-the-art in sensitivity of NMR with micro coils is around 10$^{13}$ spins/$\sqrt{\mathrm{Hz}}$ \cite{Ciobanu2002}.\\
As a result, several methods have been developed, enabling the detection of very weak magnetic resonance signals with the aim of achieving single spin detection. The three most prominent magnetometry methods, namely atomic vapor cells, superconducting quantum interference devices (SQUIDs) and magnetic resonance force microscopy (MRFM) will be briefly reviewed below. \\
Atomic vapor cells are one of the most sensitive detection methods for magnetic fields, achieving sensitivities as high as $10^{-18}\ \mathrm{T}/\sqrt{\mathrm{Hz}}$. This makes them ideal for a number of applications in NMR detection. The magnetic field measurement principle is based on the Faraday effect in a vapor of alkali atoms such as K, Rb or Cs - confined in an optically transparent cell close to the specimen to be measured. The atoms of the alkali vapor are optically polarized via a circularly polarized pump beam. In the presence of the magnetic field to be measured, generated by, for example, an ensemble of nuclear spins, the alkali atoms start to precess with a frequency corresponding to their gyromagnetic ratio. Using a linearly polarized probe beam, one can detect an NMR signal via a tilt of the probe beam polarization. The tilt is based on the Faraday rotation which is directly proportional to the strength of the external magnetic field. As a result, vapor cells are quite bulky in size. In order to maintain appropriate vapor densities and long T$_1$ times of the alkali spins one has to use vapor cells in the range of cm$^3$. A sensitivity for proton spins in a volume of around 1 mm$^3$ was estimated \cite{Ledbetter2008} to be $\sim 10^{13}/\sqrt{\mathrm{Hz}}$.\\
SQUIDs comprise a superconducting ring interrupted by two Josephson junctions, e.g. insulating barriers. Due to quantum interference only a current which is periodic in the magnetic flux penetrating the ring can flow in the ring. In d.c.\ SQUIDs, a d.c.\ current is running through the ring - this generates a voltage drop across the junction which changes in the presence of the induced currents. A change in magnetic field, caused by driving the spin transition of sample spins, can thus be directly detected electrically. SQUIDs are considered one of the most sensitive methods for detecting a magnetic field flux, reaching a noise floor of 0.33 fT/$\sqrt{\mathrm{Hz}}$ for a loop diameter of a few mm \cite{Schmelz2011}. Because the SQUID needs to be kept at cryogenic temperatures, measurements usually take place at low temperature as well. There are, however, examples for measurements with samples at room temperature \cite{McDermott2004}, although there sensitivities usually are not better than standard NMR. With nanoSQUIDs \cite{Amit2010}, on the other hand, the projected electron spin sensitivity reaches unity \cite{Vasyukov2013}, which could translate to approximately $10^3/\mathrm{Hz}^{1/2}$ for nuclear spins.\\
One of the most promising methods for magnetic resonance detection at the nanoscale is MRFM. This technique uses a magnetic tip, attached to a small mechanically resonant cantilever and placed above a spin containing sample. A nearby RF coil generates a resonant alternating field. The magnetic tip generates a magnetic field gradient, which extends into the sample volume. Just like in conventional magnetic resonance imaging (MRI) the magnetic field gradient encodes spatial information in the recorded resonance. To enhance sensitivity, experiments are carried out at cryogenic temperatures, at which the thermal fluctuations, especially of the cantilever, are greatly suppressed. In the presence of the static magnetic field $B_0$ and the inhomogeneous magnetic field generated by the magnetic tip, a specifically chosen RF frequency addresses sample spins inside a restricted sample volume (resonance slice) meeting the resonance condition. During the experiment the cantilever is oscillating above the sample while applying the alternating field. This induces periodic spin flips inside the sample which in turn exert a magnetic force onto the magnetic tip - the concomitant slight shift of the cantilever frequency is then read out. In an inverted geometry the sample is attached to the tip of the cantilever and is scanned across a static magnetic tip. In that geometry the sample is essentially moved back and forth through the resonance slice. With this detection method, sample volumes as small as (4 nm)$^3$ with an effective number of around 100 nuclear spins have been detected \cite{Degen2009}. MRFMs need to operate at cryogenic temperatures, too.\\
Technological advances in magnetometry have brought forth new magnetometers with very high sensitivity, which makes it possible to measure very small magnetic fields. On the nanoscale, magnetic fields have dipolar characteristics as they originate from single electron and nuclear spins. The strength of these fields scales like sample-detector distance, $1/r^3$. As a result, detectors on the size of the field origin itself, e.g. single atoms, are best suited to capture single electron and nuclear magnetic fields. So far, only miniaturized SQUIDs \cite{Vasyukov2013, Embon2015} and magnetic resonance force microscopes \cite{Degen2009} have exhibited few-spin magnetic field sensing with nanoscale resolution. Nevertheless, their challenging experimental requirements, with emphasis on temperature and vacuum, calls for novel sensor types. 
\section{Diamond defect spin sensors}\label{NV}
\begin{figure}[ht]
	\centering
	\includegraphics[width = 0.475\textwidth]{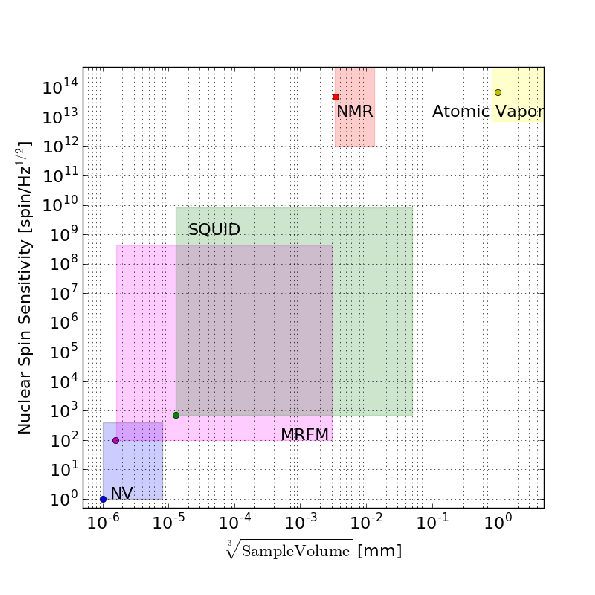}
	\label{figure1}
	\caption{Nuclear spin sensitivity for different sensing techniques plotted against their typical sensing volume: microcoil NMR \cite{Ciobanu2002}, atomic vapor \cite{Ledbetter2008}, SQUID \cite{Vasyukov2013}, MRFM \cite{Degen2009} and the NV center in diamond \cite{PhysRevLett.113.197601}. The rectangles serve as guide to the eye, depicting the space each techniques approximately covers, with the rounds markers setting the current state-of-the-art of each one.}
\end{figure}

\begin{figure*}[t]
	\centering
	\includegraphics[width=0.95\textwidth]{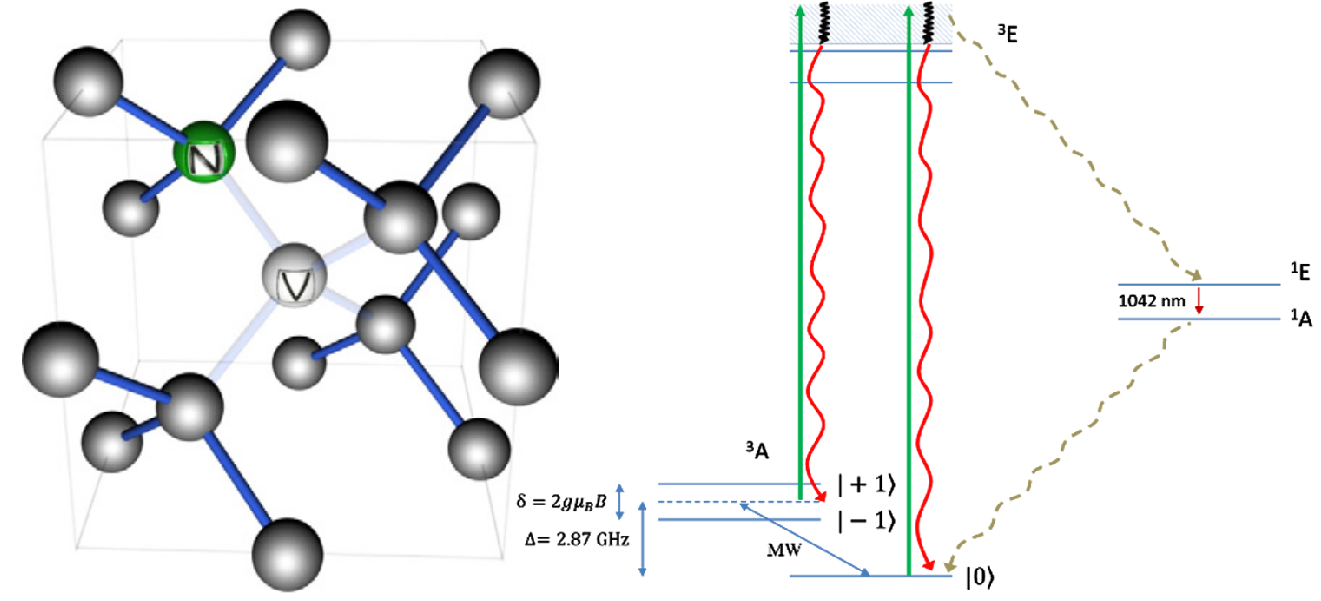}
	\caption{Left: Structure of the NV center in diamond, constituting a substitutional nitrogen (green N) and a nearest-neighbor vacancy (white V). The other gray atoms represent the carbon in the diamond lattice. Reprinted from Ref.~\citealp{Wrachtrup2006}. Right: The energy level diagram for the NV center. The triplet ground state and the excited state are separated by 1.945 eV, which is $\sim 638$ nm. At zero magnetic field, the ground state $\ket{m_s=0}$ is separated by 2.87 GHz from the $\ket{m_s=\pm1}$. Off-resonant excitation at 532 nm can result in a spin-conserving radiative decay at 638 nm or through a inter-system crossing from triplet to singlet resulting in a lower fluorescence rate at 638 nm - depending on the initial ground spin state.}
	\label{figure2}
\end{figure*}

\begin{figure*}[ht]
	\centering
	\includegraphics[width = 1.00\textwidth]{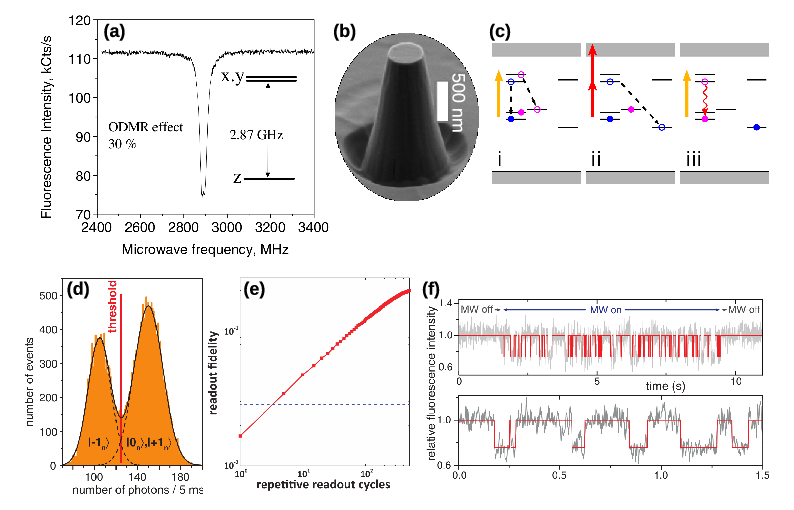}
	\caption{(a) Fluorescence-detected magnetic resonance of a single NV center in diamond. This spectrum was measured at ambient conditions at zero external magnetic field. Reprinted from Ref.~\citealp{Jelezko2006}; (b) A tapered, nanoengineered diamond waveguide, increasing photon counts from the NV center fivefold. Reprinted from Ref.~\citealp{Momenzadeh2015}; (c) Spin-to-charge conversion level diagram scheme. Instead of measuring the spin state of the NV, the idea here is to transfer it to a charge distribution and then read it out. Blue circles indicate the $\ket{m_s=0}$ spin state, pink - $\ket{m_s=\pm1}$. Yellow and red arrows are orange (594 nm) and red (638 nm) laser excitations, respectively. Taken from Ref.~\citealp{Shields2015}; (d) Photon-counting histogram of a fluorescence time-trace, fitted by two Gaussian distributions. The left peak corresponds to the $\ket{-1_n}$ nuclear spin state of the $^{14}\mathrm{N}$ nucleus in the NV center, and the right peak to the $\ket{0_n},\ket{+1_n}$ states. Setting a threshold (red line) enables one to resolve the left state from the right other states. Reprinted from Ref.~\citealp{Neumann2010b}; (e) The achievable gain in readout fidelity as a function of number of repetitive readouts. The blue line indicates the fidelity achieved by standard readout. Taken from Ref.~\citealp{Lovchinsky2016}; (f) Fluorescence time trace showing quantum jumps of a single nuclear spin. When MW pulses (Single-shot readout with CNOT gates) are on, the signal starts switching between the two states from (a). Reprinted from Ref.~\citealp{Neumann2010b}.}
	\label{figure3}
\end{figure*}

Sensing magnetic field with spins is a prevalent technique. Sensitive NMR probes to measure fields are common techniques for a number of applications including field stabilization of NMR and EPR spectrometers themselves. Typically, the accuracy reached is in the order of a few parts per million (ppm) with fields ranging up to 20 T, i.e.\ 10 \si{\micro\tesla}/$\sqrt{\mathrm{Hz}}$. This is equivalent to the field of a single electron spin at a distance of 1 \si{\micro\meter}. Despite their reasonable sensitivity, such spin probes are not sensitive enough to measure single electron spins since their size is on the order of a few cm$^3$. As a result, spin sensors need to be small without compromising their sensitivity for single spin detection. Ultimately, such sensors contain a single spin themselves and can be brought into close proximity to the sample. The sensitivities of the different techniques presented in Sec.~\ref{intro} are plotted in Fig.~\ref{figure1}.
In the past, several systems have been identified with the potential of detecting single spins in solids. Nearly all of those have been obtained via a combination of optical measurement and spin resonance. Specifically the first such optically detected magnetic resonance experiments were done on single molecules at low temperature, basically combining single molecule spectroscopy and electron spin resonance \cite{Wrachtrup1993a}. These early experiments relied on the existence of an excited metastable triplet state, the lifetime of which depends on the specific spin state. Spin Rabi oscillations as well as spin coherence was measured \cite{Wrachtrup1993a, Wrachtrup1993b, Wrachtrup1995} and for the first time ideas of using single molecules as sensitive and highly local magnetic field probes were conceived \cite{WrachtrupApplication1995}. Yet, as the spin carrying state is an excited state with limited lifetime, sensitivity as well as versatility is greatly reduced. Few systems are known, which have a paramagnetic ground state and at the same time show strongly allowed optical transitions. Among those are quantum dots \cite{Berezovsky2006} and defect centers in solids \cite{Gruber1997, KolesovXiaReuterEtAl2012}. 
Certain materials, most notably diamond and silicon carbide, combine two outstanding properties: On the one hand they are wide band-gap semiconductors (band-gap 5.48 eV  for diamond and 3.5 eV for silicon carbide, respectively) and on the other hand both materials show a very weak spin-orbit coupling. The first property allows single defects to be addressed optically in convenient wavelength ranges. The latter ensures long spin relaxation times, a prerequisite for sensitive detectors. The sensor principle shall be explained on the nitrogen vacancy (NV) center in diamond as this is one of the best studied systems. 
The NV center is one of the most prominent color centers in diamond. It comprises a nitrogen substituting a carbon atom and a nearest neighbor vacancy (see Fig.~\ref{figure2}).
The electronic wave function is a linear combination of three electrons in the carbon dangling $p$-orbitals, one electron from the nitrogen lone pair and one electron from donors in the lattice. The electronic ground state of the defect is an electron spin triplet ($S=1$). Optical excitation of the defect commences to a spin triplet state as well with optical transitions conserving spin orientation to first order. Most importantly, however, the defect shows strong spin polarization upon optical illumination. This comes as a result of a metastable electronic state which is energetically positioned between the excited triplet state and the ground state. Instead of relaxing to the ground state the excited electron can relax via this metastable state. The matrix element for this transition is determined by spin-orbit coupling, which is a highly spin-state selective process. As the lifetime of this state is on the order of 300 ns, 30 times longer than the lifetime of the excited singlet state, the resulting fluorescence intensity is significantly lower. The principle is known as fluorescence detected magnetic resonance and has been applied to a number of systems in the past, for example organic molecules. Fig.~\ref{figure3}a shows a fluorescence detected electron spin resonance spectrum of a single NV center. It is characterized by a field independent transition frequency, the zero field splitting which is a characteristic of every electron spin system with $S \geq 1$.
The Hamiltonian describing the system is given as
\begin{equation}
	H = S\bar{D}S+g\beta B_0S + S\bar{A}I,
\end{equation}
where $\bar{D}$ is the zero field splitting tensor, $B_0$ is the external magnetic field and $\bar{A}$ is the hyperfine coupling tensor. The resonance signal shows the well-known Zeeman shift. This is the basis for using the NV as a magnetic field sensor - the capability of measuring local EPR and NMR signals. As with other magnetic field sensors the sensitivity of this sensor is of prime interest for the present purpose. This accuracy or sensitivity is given by $\eta=\dfrac{\pi\hbar}{2g\mu_BC\sqrt{T_2}}$ \cite{Taylor2008}. As can be seen from the equation, $\eta$ depends on the spin dephasing time $T_2$. For NV center spins, this dephasing time is limited by hyperfine coupling between the electron and nuclear spins of $^{13}\mathrm{C}$ in the diamond lattice. For a natural concentration of $^{13}\mathrm{C}$ of 1.1\% it is around 0.6 ms \cite{Markham2011}. Reducing $^{13}\mathrm{C}$ down to levels below 0.01\% has yielded $T_2$ values of up to 2.0 ms \cite{Yamamoto2013}. This results in an $\eta$ of around 4 nT with typical values for $C$, the signal to noise ratio for the NV electron spin detection. This is the magnetic field of an electron spin at a distance of around 1 \si{\micro\meter} or of a nuclear spin at 10 nm. The spin wavefunction of the defect is almost exclusively localized on the dangling bonds of the three carbon atoms comprising the defect. As a result, the defect can be brought close to surfaces without the wavefunction and the optical- or spin properties fundamentally compromised. This is the basis for ultrasensitive measurement of single electron or nuclear spins using the defect center.
\section{Reading out the spin sensor}
In most applications the spin defect magnetic resonance is measured by detecting the fluorescence intensity of the defect center. In room temperature applications, where most recent studies have been concentrating their efforts due to its convenient features, the fluorescence is excited by a 532 nm laser which has been shown to be optimal for generating fluorescence of the negative charge state of the defect. The magnetic resonance signal of the defect ground state is detected as a change in the fluorescence intensity as already described in Sec.~\ref{NV}. Typically, this yields a decrease of about 30\% in fluorescence intensity. Alternatively, readout of the spin signal at low temperatures ($T<10$ K) can be done by spin state selective optical excitation. While low temperature applications of the defect are currently more exotic, the mechanism should be described nevertheless as it points to an interesting feature to enhance room-temperature readout: At low-temperature the optical excitation lines are narrow enough to address individual spin sublevels separately. As a result, no microwaves are necessary to measure Zeeman shifts if the ground and excited states shift differently. More importantly, the spin measurement can now be digitized and converted to a spin state measurement. To understand the difference over a conventional spin state it is important to consider the following case: If the time it takes to measure a certain spin quantum state is smaller than the spin lattice relaxation time one can determine the spin state in addition to the conventional spin transition frequency. This situation occurs at low temperature - here the spin lattice relaxation time of the defect center electron spin approaches seconds. On the other hand, the fluorescence rate is around $10^5$ photon counts/s. In other words, it is possible to detect $10^5$ photons before a spin flip occurs. One can now define a threshold for the fluorescence counts to decide in which spin state it is in. Such an approach is called single shot readout (SSR) and could considerably enhance signal-to-noise in experiments \cite{Robledo2011}. This quantum logic readout cannot be transferred to room temperature in a straightforward fashion as the spin relaxation time is reduced to some ms. As a result, the average number of photons per spin state is $\langle n\rangle \sim 0.1$ for $m_s=0$ and $\langle n\rangle \sim 0.07$ for $m_s=\pm 1$ and the standard deviation of the two distributions ( $\sqrt{\langle n \rangle}$ ) of the two distributions overlap considerably. Consequently, a state discriminating readout is not possible. Nevertheless, it turns out that this is not the case for nuclear spin states. For nuclear spins having the proper orientation, i.e.\ hyperfine coupling with the electron spins, their spin state is unperturbed for a certain number of readout cycles, $M$. Therefore, single shot readout becomes feasible (see Fig.~\ref{figure3}b-d).  Through a sequence of quantum gates, the electronic spin state $\alpha\ket{0}+\beta\ket{1}$ (which contains information about the external magnetic field) can be correlated with the nuclear spin, resulting in the entangled state $\alpha\ket{0\downarrow}+\beta\ket{1\uparrow}$. Here $\ket{\downarrow}$,$\ket{\uparrow}$ denote the nuclear spin quantum states. An optical measurement then projects the state into either $\ket{0\downarrow}$ or $\ket{1\uparrow}$ and repolarizes the electron spin to $\ket{0}$. Even when the electron spin is reset, the nuclear spin retains information about the initial populations\footnote{Relatively high magnetic fields ($> 0.4 T$ are essential here in order to have decoupling between the nuclear and the electron spin dynamics.}. We can then apply a controlled-NOT gate (CNOT) to map the state, e.g.\ $\ket{0\downarrow} \rightarrow \ket{0\downarrow}$ and $\ket{0\uparrow} \rightarrow \ket{1\uparrow}$ and perform another optical measurement. This process can be repeated multiple times, thus collecting more information about the original electronic spin state. These CNOT gates (and the initial mapping of the electron spin state to the nuclear spin state) require a finite time in the measurement sequence compared to classical measurement schemes, and so it is worth defining the overall readout noise and compare it to the single shot readout. In the case of a classical scheme the overall signal-to-noise-ratio (SNR) is $\mathrm{SNR}_\mathrm{classical} = \left(N_{\ket{0}} - N_{\ket{1}}\right)/\sqrt{N_{\ket{0}} + N_{\ket{1}} + N_b}$. Here $N_{\ket{0},\ket{1}}$ is the number of photons for the spin being in spin state $m_s=0$ or $\pm 1$ and $N_b$ is the number of background fluorescence photons. For single shot readout, the SNR depends on the fidelity, $\mathcal{F}$, i.e. the ability to distinguish\footnote{A fidelity of 1.0 means that there is a 100\% probability to distinguish between the two, and so any (realistic) fidelity smaller than 1.0 reduces the SNR.} whether the nuclear spin is $\ket{\uparrow}$ or $\ket{\downarrow}$, such that $\mathrm{SNR}_\mathrm{SSR} = \sqrt{M\mathcal{F}p(\ket{\uparrow})p(\ket{\downarrow})}/p(\ket{\uparrow})$, where $p(\ket{\uparrow})$  is the probability to measure the nuclear spin at $\ket{\uparrow}$. The signal improves with the square root of the number of repetitions, $\sqrt{M}$ (see Fig.~\ref{figure3}e). Typically, improvements on the order of a factor of 3-5 can be achieved, and are highly dependent on the preceding sensing period and spin-state evolution. Another promising scheme is to map the NV's spin state onto a charge distribution, which is then read out optically \cite{Shields2015}. This technique shows a potential for a threefold reduction in readout noise (see Fig.~\ref{figure3}c for a level diagram sketch of this method). A further decisive number is the total signal, i.e. fluorescence intensity $N_{\ket{0}}$ or $N_{\ket{1}}$ as the noise scales with $\sqrt{N_{\ket{0},\ket{1}}}$. As a result, in most experiments great care is taken to collect as many photons as possible from the NV center. Mainly this is done via excellent (commercial) collection optics and diamond waveguides. Simple examples are tapered diamond rods \cite{Momenzadeh2015} at the top of which the defect is situated (see Fig.~\ref{figure3}b). With such structures SNRs for the detection of 1 defect on the order of 1000 can be achieved with averaging times of 1 s. Such signal intensities are crucial for the use of NV centers as detectors as they directly enter into the sensitivity of the sensor. 
\section{Detecting magnetic fields}\label{detect}
Diamond defect centers have been shown to detect a variety of parameters. These are temperature \cite{Neumann2013}, force \cite{Doherty2014} or electric field \cite{PhysRevLett.112.097603}. The most natural quantity to detect, however, is magnetic field. As already discussed, this is done by accurately measuring the Zeeman shift of the defect electron spins. A limiting parameter is the dephasing time of the defect spin $T_2$, which can reach $T_1$, the latter being on the order of 2-5 ms at ($T = 300\ \mathrm{K}$), depending on the sample's quality. However, as typical for most solid state systems, $T_2^*$ is significantly shorter - on the order of a few \si{\micro\second} to one hundred \si{\micro\second}. As a result, high sensitivities are only achieved when refocussing pulses or more generally if decoupling sequences are applied to the sensor spin. These sequences, which resemble in many ways classic NMR and EPR ones, limit the bandwidth of the sensor, rendering it sensitive only above a certain frequency, which is on the order of a few kHz. In demonstration of sensitive measurements of magnetic fields (and this is the case for all sensing applications) a decoupling sequence is run on the NV spin while the field to be measured is modulated in synchrony with a fixed phase to the sequence. Using this technique, sensitivities as high as 0.9 pT/$\sqrt{\mathrm{Hz}}$ have been achieved with ensembles of around $10^{11}$ defects with measuring fields as small as 100 fT \cite{Wolf2014}. 
The same techniques, enabling the detection of weak magnetic fields, are also used to detect single electron or nuclear spins. At this nanoscale regime, statistical polarization\footnote{The difference in numbers between spin-up and spin down. This difference fluctuates according to random flips of the spins} can overcome the thermal, or Boltzmann polarization. This happens when $V<\left( \mu B/k_BT\right)^{-2}\rho_N^{-1}$, where $\mu$ is the spin's magnetic moment, $B$ is the magnetic field and $\rho_N$ is the spin number density \cite{Degen2007}. 
There are different ways to a analyze those measurements, depending on whether the external field modulation (i.e. the sample spins' magnetic field) occurs in a phase coherent, that is, controlled way, or if the modulation is a random process, i.e. resembling a classical noise source. Since current applications mostly focus on room temperature sensing, the latter approach should be highlighted first. The fluctuation of spins is most conveniently described as classical magnetic field noise, which reduces the defect center's (sensor) spin coherence. In a semi-classical approach this is described as the average over a randomly fluctuating phase $\varphi$,
\begin{equation}
	C(t) = \left\langle e^{i\varphi( t )} \right\rangle = e^{\chi(t)}.	
\end{equation}
Here $\chi(t) = \int \mathrm{d}\omega\left[\dfrac{S(\omega)}{\omega^2}\right]F(\omega; t)$ with $S(\omega)$ being the noise power spectral density of the magnetic field fluctuations and $F(\omega;t)$ being the filter function of the applied pulse sequence. $F(\omega;t)$ relates to the pulse sequence by a Fourier transformation. At the NV transition itself one can employ a spin relaxation measurement with an effective narrow Lorentzian $F(\omega)$ around the transition frequency. A two-pulse echo sequence (Hahn Echo), on the other hand, has an $F(\omega)$ with a rather large bandwidth whereas multipulse sequences like CPMG \cite{Cywinski2008} or more robust sequences like XY8 \cite{deLange01102010} have a narrower one and as a result are more suited to reject a wider noise bandwidth.
\begin{figure*}[ht]
	\centering
	\fbox{\includegraphics[width = 0.75\textwidth]{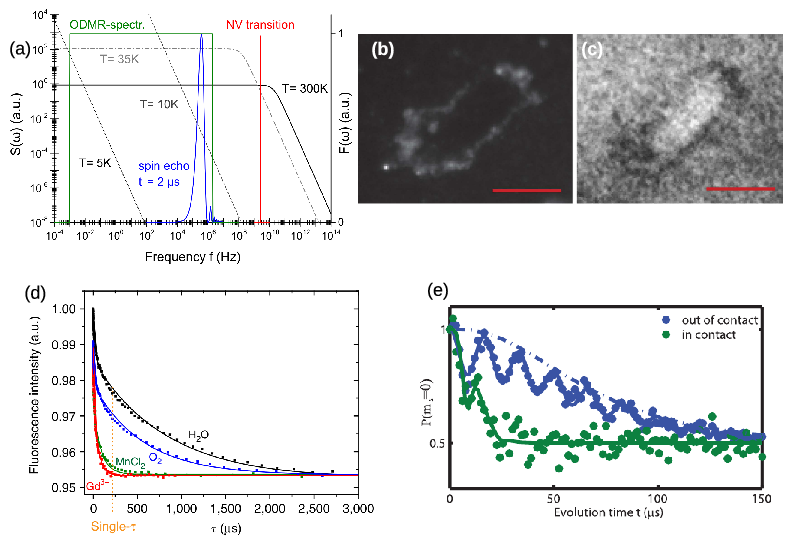}}
	\caption{Detecting magnetic fields at a wide frequency range: (a) The spectral sensitivity $F(\omega)$ for different magnetic field sensing filter schemes using the NV center in diamond: $T_1$ relaxometry (red), $T_2$ dephasing or Hahn Echo (blue) and ODMR spectroscopy (green). The dashed lines are overlays of the spectral density function, $S(\omega)$, of ferritin molecules; (b) Fluorescence image of a slice of HeLa cell (150 nm), in which its membrane was labeled with biotin-poly-L-lysine-Gd$^{3+}$-DTPA-Alexa532; (c) Magnetic imaging ($T_1$-relaxation) performed simultaneously with (b), giving evidence for the presence of Gd$^{3+}$ at the membrane of the cell; (d) Effect of different electron spin species on the relaxation time, $T_1$, of an ensemble of NVs. The largest effect is from Gd$^{3+}$ spins; (e) Effect of surface spins on the dephasing time, $T_2$, of a single NV - the dephasing is much faster when the scanning NV is in contact with the surface. Sources: (a): Ref.~\citealp{Eike2014-2}, (b)-(d): Ref.~\citealp{Steinert2013}, (e): Ref.~\citealp{Luan2015}.}
	\label{figure4}
\end{figure*}
A key feature is that the maximum has an amplitude which scales with $N^2$ when $N$ is the number of pulses, whereas the width of the transmission peaks scales as $\sim t^2$, when $t$ is given by $\omega t = (2k-1)N\pi$, where $k$ is the $k^\mathrm{th}$ pulse out of $N$ pulses in total. By changing the pulse spacing, i.e. changing the number of pulses $N$ for a given $t$, one can scan the whole noise spectrum and single out a specific component. This is the basis of one of the techniques used to measure NMR signals with the NV center (see Sec.~\ref{NMR}). Fig.~\ref{figure4}a shows a schematic representation of different filter functions used by the NV center in diamond to detect magnetic fields at several frequency regimes. For each scheme, representative measurements are shown (Figs.~\ref{figure4}b-e).

\section{Detecting electron spins}
Initial experiments to detect spins with the NV center have concentrated on detecting electron spins. Though electron spins have a large magnetic moment and can be sensed by the NV center at more than 10 nm distance, their detection by the NV center has remained elusive for quite some time. Mostly this is due to the presence of a wealth of electron spins on the surface of the diamond which themselves contribute to the signal. In preliminary experiments, dipolar coupling between the NV and unknown paramagnetic species on the diamond surface was detected \cite{Grotz2011}. Shallow NVs with distances between 3 and 10 nm below a diamond surface were used to detect electron spins on the surface. The surface was silanized and 4-Maleimido-(2,2,6,6-Tetramethylpiperidin-1-yl)oxyl (TEMPO) radicals were attached to it. A standard double electron-electron resonance (DEER) sequence was used to identify coupling with a single electron spin - a coupling strength of 600 kHz corresponding to a distance of 3.5 nm. In a more systematic study double resonance studies with the detection of 2,2-diphenyl-1-picrylhydrazyl (DPPH) electron spins have been carried out. As in the first case, a DEER sequence was used to detect ensembles of electron spins on the diamond surface using a single NV spin roughly 15 nm below the surface \cite{Mamin2012}.
A whole set of intriguing imaging experiments of small ensembles or single electron spin using single NV centers have been carried out since then: Coupling to a single defect spin within diamond was detected \cite{Shi2013} and the distance as well as relative orientation was measured precisely \cite{Grinolds2014}. While there are cumulative examples for detecting electron spins inside and on the diamond surface, the detection of electron spins in samples outside the diamond has been relatively scarce so far. The main reason is that there is a sizeable density of electron spins around the NV center itself and specifically on the surface (0.5/nm$^2$) which easily masks the electron spins in the sample to be detected.

\begin{figure*}[ht]
	\centering
	\includegraphics[width = 0.95\textwidth]{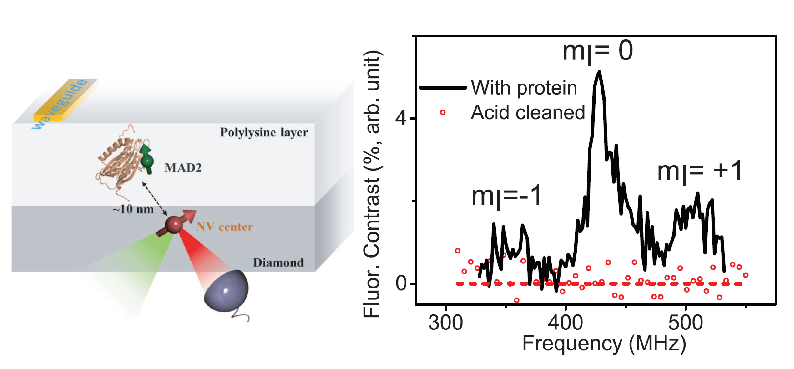}
	\caption{Left: The spin label is a cysteine-MTSSL on a monomer mitotic arrest deficient-2 (MAD2) protein, located on the surface of the diamond, near an implanted NV center. Right: Single spin EPR spectra under ambient conditions. The origin of the three peaks is the nitroxide's spin label hyperfine coupling to its $^{14}\mathrm{N}$ nucleus ($I=1$). The spectrum disappears after removing the protein by acid cleaning. Reprinted from Ref.~\citealp{Shi2015}.}
	\label{figure5}
\end{figure*}

Recently, the DEER spectrum of a nitroxide spin label attached to a single MAD2 protein was measured by a single NV center implanted approximately 5 nm below the surface of the diamond \cite{Shi2015}. The spectra, shown in Fig.~\ref{figure5}, exhibit three distinct peaks, which are a mark of the spin label's $^{14}\mathrm{N}$ nuclear spin ($I=1$) hyperfine interaction with its electron spin ($S=\frac{1}{2}$).
This is in fact the first work, which shows the actual sensing of a single electron spin in a biomolecule by the NV center in diamond with a unique spectral signature. 
\section{Detecting nuclear spins}\label{NMR}
\begin{figure*}[ht]
	\centering
	\includegraphics[width = 0.75\textwidth]{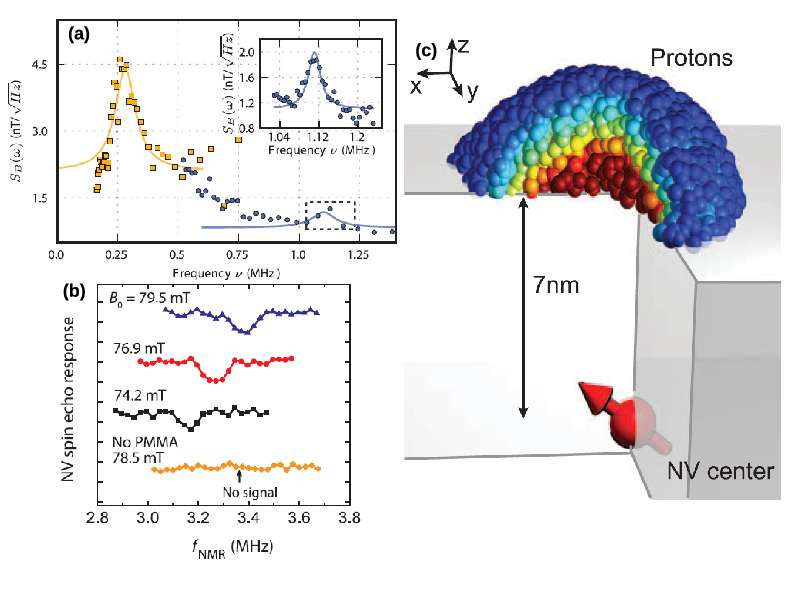}
	\caption{First demonstration of nanoscale $^1\mathrm{H}$-NMR detection. (a) $^{13}\mathrm{C}$ signal from the nuclei inside the diamond (yellow) and $^1\mathrm{H}$ signal from microscopy immersion oil. Taken from Ref.~\citealp{Staudacher2013}; (b) Right: $^1\mathrm{H}$ signal from PMMA. The peak position shifts with magnetic field. Taken from Ref.~\citealp{Mamin2013}; (c) Numerical simulation of the detection volume in (a) - a three-dimensional visualization of the $10^4$ closest protons, generating 70\% of the signal (shown as spheres). The color signifies the magnetic field the protons contribute to the signal, with red = 4 nT and blue = 1.5 nT.}
	\label{figure6}	
\end{figure*}

\begin{figure*}[ht]
	\centering
	\includegraphics[width = 0.95\textwidth]{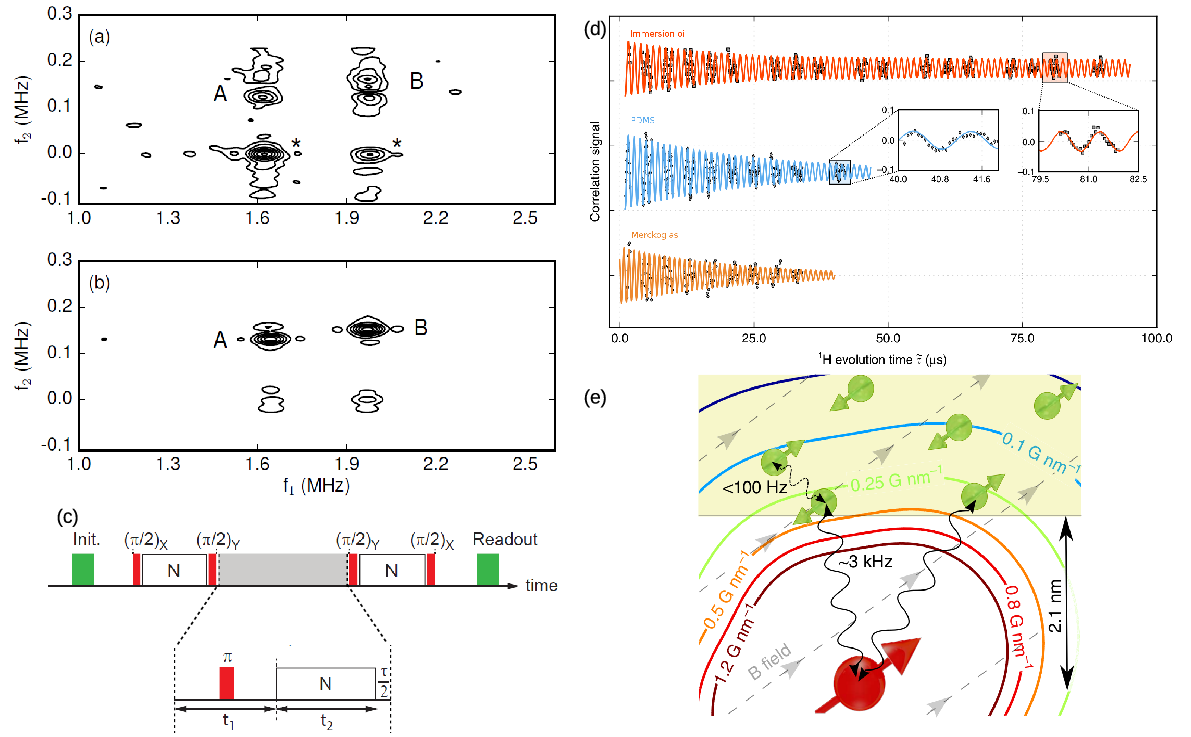}
    \caption{Two-dimensional homonuclear ($^{13}\mathrm{C}$) NMR spectrum, measured by spin-correlation spectroscopy with an NV center in diamond. The image shows the 2D-FFT of the acquired data. $f_1$ is the nucleus's bare Larmor frequency plus its parallel hyperfine coupling to the NV, while $f_2$ is its perpendicular hyperfine coupling to the NV. (a) experiment and; (b) simulation; (c) The pulse sequence used to measure (a) and simulate (b). Reproduced from Ref.~\citealp{Boss2015}; (d) Correlation spectroscopy signals from three different samples, exhibiting different diffusion times (correlation time). Taken from Ref.~\citealp{Staudacher2015}; (e) Sensing nuclear spins in the strong-coupling regime, where the signal is proportional to the number of spins. The countour lines represent the effective magnetic field from the NV (red) felt by the target $^{29}$Si nuclei (green) due to hyperfine interaction. Reproduced from Ref.~\citealp{Mueller2014}.} 
   	\label{figure7}
\end{figure*}

While sensing few electron spins has been already presented in recent works, the detection of nuclear spins has been mostly limited to an ensemble of approximately 100 nuclei of the same type \cite{Mamin2009}, mainly due to the difference between the electron's and the nucleus's gyromagnetic ratio, which determines the coupling strength between the NV's electron spin and the target spin. To increase this coupling, two routes have been pursued: (a) Attempting to have the NV as shallow as possible, i.e. as close as possible to the external target spin, and (b) increasing the NV's magnetic field sensitivity. Typically, all proof-of-concept experiments with NVs for the detection of nuclei were performed on the diamond host matrix's $^{13}\mathrm{C}$ ($S=\frac{1}{2}$), which have a natural abundance of 1.1\% in diamond. These have the convenience of being as close as one could possibly have a nucleus to the NV. Indeed, several standard-setting examples have exhibited the detection and coherent manipulation \cite{Laraoui2010, Taminiau2012, Laraoui2013} of carbon nuclei in diamond. In these experiments, a Hahn echo pulse scheme was used to decouple the NV from noise other than the target nuclei spin species.
As already stated in Sec.~\ref{detect}, when using more advanced dynamic decoupling pulse sequences or by actively flipping the nuclear spin using an RF pulse, this aforementioned filter function can be narrowed down in frequency, thus making it possible to distinguish between different nuclear species. Two seminal works used such pulse sequences to detect the signal from a proton ensemble in immersion oil \cite{Staudacher2013} and poly-methyl-methyl-acrylate (PMMA) \cite{Mamin2013}. Fig.~\ref{figure6} shows a proton signal from these measurements.
In both cases, the NV was located only several nm below the surface of the diamond, $d$, and correspondingly the relevant sensing volume from which 70\% of the nuclei were contributing signal was $d^3$. For such small volumes, the number of nuclei having a magnetic field effect on the NV is on the order of $N = 10^4$. Although the mean magnetization of these spins would be zero, there would be a significant fraction, $\sqrt{N} = 10^2$, which would be statistically polarized along a random direction \cite{Degen2007}. In the case of the first method of actively driving the nuclear spins \cite{Meriles2010, Mamin2013}, this fraction of statistical polarization results in the NV being coupled to the longitudinal magnetization of the nuclei, $\langle M_z \rangle$. Conversely, in the case of passively sensing the nuclei's Larmor precession \cite{deLange01102010, Staudacher2013}, the NV is sensitive to the transverse magnetization, which is being the generator of the signal, with $B_z = B\left(\langle M_x \rangle, \langle M_y \rangle \right)\cos\left(2\pi t / \tau_L + \phi\right)$, where $\tau_L$ is the Larmor period of the nuclei. 
Nevertheless, the latter techniques still suffer from a relatively broad bandwidth of several kHz\footnote{This is highly dependent on the distance of the NV center from the surface of the diamond. ``Deep'' NVs offer a long coherence time and a high-order dynamic decoupling sequence can be used, but of course they are not as practical for NMR purposes. ``Shallow'' NVs, i.e. ones which are a few nanometers below the surface, have a shorter coherence time, and as stated above, a linewidth of a few kHz using dynamic decoupling schemes.}, which, as was clearly pointed out \cite{PhysRevX.5.021009}, can lead to inconclusive determination of the nucleus's identity. One of the limitations of these dynamic decoupling (DD) schemes is the dependence on the NV spin's decoherence time, $T_2$. This prevents the use of long waiting times, $t$, between adjacent $\pi$-pulses, which in turn limits the maximum number of pulses. Yet another hardware limitation is the ability to produce ever shorter pulses (for achieving the same $\pi$ flip of the NV but in a shorter time, one needs to increase the power of microwave power, or transverse magnetic field). Nevertheless, with the application of repetitive readout \cite{Jiang2009} and the use of highly-purified materials, it is now also possible to detect nuclei from a single protein  \cite{Lovchinsky2016}.
This calls, then, for yet more complex or sophisticated schemes, which overcome or avoid altogether these barriers. One of the most fundamental building blocks in this respect is spin-correlation spectroscopy, in which two DD pulse sequences are spaced apart by a free-evolution time, $T_C$. In each of the DD blocks, the NV acquires a phase, $\varphi_1$ and $\varphi_2$, and then for different $T_C$, the correlation between these two phases, $\left(\langle \sin\left(\varphi_1\right)\sin\left(\varphi_2\right)\right\rangle$, is in effect measured. If $T_C$ is a multiple of the target nucleus Larmor period, the phases measured in each of the DD blocks will be the same, which is essentially a positive correlation. In the other extreme case, wherein $T_C$'s length is longer by one target nuclear spin half Larmor period, the phases are opposite and the correlation would be negative. This would lead to a periodic oscillation, limited by the NV's relaxation time, $T_1$, with possible decay due to interactions of the nuclei with neighboring nuclei or with surface dipoles \cite{Staudacher2015}.
A display of this technique has been initially demonstrated on $^{13}\mathrm{C}$ spins \cite{Laraoui2013} and $^1\mathrm{H}$ spins \cite{Kong2015b}, and more recently as a tool to probe molecular dynamics \cite{Staudacher2015} (see Fig.~\ref{figure7}d) and achieve chemical contrast, and nuclei hyperfine couplings \cite{Boss2015}. Spin correlation spectroscopy thereby overcomes the $T_2$ limitation, and as stated above, bounded only by the NV's own spin relaxation time, $T_1$. The NV's spin correlation spectroscopy enables measurement schemes which are very similar to conventional two-dimensional NMR, whether homonuclear or heteronuclear \cite{Aue1976}. A striking example is shown in Figs.~\ref{figure7}a-c.
In this experiment, the authors used a spin-correlation spectroscopy sequence, which consists of two components, one to probe the natural Larmor frequency of the nuclei plus its parallel hyperfine coupling to the NV center, $\omega_L + a_\parallel$, while the other probes the perpendicular hyperfine coupling to the NV, $a_\perp$. Similar to conventional 2D-NMR, the different pixels in the image are the result of running the measurement for different pulse sequences during the free evolution time, while keeping the DD blocks fixed. As one moves from a large spin ensemble to the statistical polarization regime, one is approaching the single spin NMR limit, where the strong-coupling regime persists. Here the coupling between the sensor and the target nucleus is larger than that between neighboring target nuclear spins. In this regime, which is also beyond the statistical polarization limit, the signal is now proportional to the number of individual spins (see Fig.~\ref{figure7}e), and one can then experimentally measure the signal dependence on the distance between the sensor and the target spin, verifying the strong-coupling nature of the interaction \cite{Mueller2014}.
\noindent This leads us, then, to a point where we can finally outline a comparison between conventional NMR, the three alternative techniques introduced in Sec.~\ref{intro}, and the NV center in diamond. We summarize the five different techniques in Table\ \ref{table1}, citing their sensing volume and nuclear spin sensitivity.
\begin{table*}[ht]
	\centering
	\caption{Comparison of different nuclear spin sensing techniques, from the conventional (microcoil) NMR through atomic vapor, SQUIDs, MRFM and finally the projected sensitivity of the NV center in diamond.}	
	\label{table1}
	\begin{threeparttable}	
	\begin{tabular}{lrcc}
	\toprule
		Technique	&	Sensing volume	&	Sensitivity [spin/Hz$^{1/2}$]	&	Reference						\\
		\midrule							\\
		Conv.\ NMR  &   40 \si{\micro\meter\cubed}        &   $4.9\cdot 10^{13}$             	&	\cite{Ciobanu2002}				\\
		Atomic vapor\tnote{1}& 1 $\mathrm{mm}^3$	&	$7\cdot 10^{13}$							&	\cite{Ledbetter2008}	\\
		SQUID\tnote{2}		& 2100 $\mathrm{nm}^2$	&	$700$							&	\cite{Awschalom1988, Vasyukov2013}	\\	
		MRFM		& 64 $\mathrm{nm}^3$&	$100$								&	\cite{Degen2009}	\\
		NV\tnote{3}			& 1 $\mathrm{nm}^3$	&	$1$									&	\cite{Mueller2014, PhysRevLett.113.197601}\\
		\bottomrule
	\end{tabular}
	\begin{tablenotes}
		\footnotesize
		\item $^1$ SNR 1
		\item $^2$ In SQUIDs typically the flux sensitivity, $\Phi_n$, is in units of quantized flux, i.e. $\Phi_0/\mathrm{Hz}^{1/2}$, where $\Phi_0 = h/2e \approx 2.07\cdot 10^{-15}\ \mathrm{T}\cdot \mathrm{m}^2$. The projected electron spin sensitivity is given by $ S_n = (R/r_e)\Phi_n$, where $R$ is the SQUID loop radius and $r_e$ is the classical electron radius. Here we project to nuclear spin sensitivity by multiplying by the ratio of $\mu_n/\mu_B \approx 1835$.	
		\item $^3$ Not per $\sqrt{\mathrm{Hz}}$ strictly speaking. The key point here is the volume and the ability to confirm sensing of a single spin. See also Sec.~\ref{singlespin}.
	\end{tablenotes}	
	\end{threeparttable}
\end{table*}

This already puts the NV center in diamond in the forefront of single nuclear spin sensing. As we will show in Sec.~\ref{summary}, there is still substantial room for improvement. 
\section{Nano Magnetic Resonance Imaging}\label{nanoMRI}
Magnetic resonance using single defects in diamonds achieves its most illustrious aspect on the one hand and most practical one on the other hand, when using it for imaging. In a pioneering work, a single NV center has been used to image a single NV center's electron spin. The experiment used an NV spin on a scanning nanopillar cantilever as a magnetometer with a d.c.\ sensitivity of around 2 \si{\micro\tesla}/Hz$^{1/2}$ and an a.c.\ sensitivity of 56 nT/Hz$^{1/2}$ down to 18 nT/Hz$^{1/2}$. Such a sensitivity is sufficient to detect a single NV electron spin at a distance of around 50 nm, which corresponds to a magnetic field of roughly 10 nT. To achieve such sensitivity a robust dynamic decoupling sequence (XY8) was applied to the sensor NV spin. In synchrony with the $\pi$ pulses of this sequence the sample NV spin was inverted - and indeed, a single NV spin at a vertical distance of around 50 nm was detected and imaged. In the demonstrated magnetic field imaging, single spin measurements with a signal-to-noise ratio of one could be acquired in around 2 min \cite{Grinolds2013}.
\begin{figure*}[ht]
	\centering
	\includegraphics[width = 0.75\textwidth]{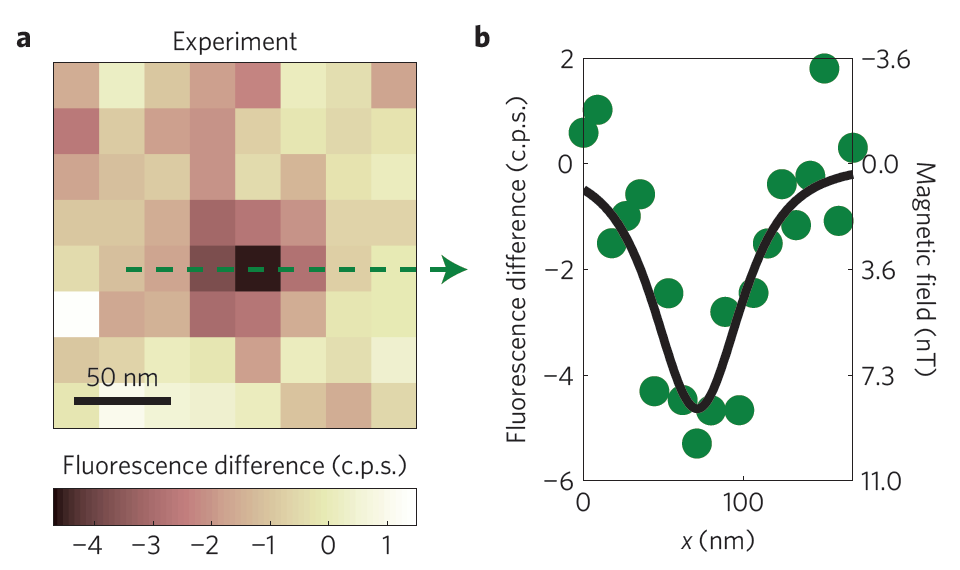}
	\caption{Magnetic field image of a target NV spin near the surface of a diamond mesa, acquired with a scanning NV magnetometer. While repeatedly running an a.c.\ magnetometry pulse sequence (here with a 32-pulse XY8 sequence, with 40 \si{\micro\second} of total evolution time), the sensor NV is laterally scanned over the target, and the fluorescence rates for the target spin starting in the $\ket{0}$ state as well as the $\ket{-1}$ state are independently recorded. Plotted is the difference between these measurements, which depends only on the sensor NV’s magnetic interaction with the target spin and not on background fluorescence variations. The pronounced drop in fluorescence near the centre of the image indicates a detected single electron spin. Reprinted from Ref.~\citealp{Grinolds2013}.}
	\label{figure8}
\end{figure*}
The data shown in Fig.~\ref{figure8} have been averaged for 40 min per point, yielding an SNR of 4, which is a 15-fold reduction as compared to MRFM measurements reporting single electron spin imaging.  Similarly, single and multiple “dark” electron spins within diamond have been imaged by the scanning field gradient method. More specifically DEER has been used to address $g=2$ electron spins at or very close to the surface of diamond. Using scanning tips with field gradients between 2.4 and 12 G/nm the authors were able to demonstrate that the dark spins are located around 10-14 nm separated from the NV center \cite{Grinolds2014}, with a coupling strength of approximately 100 kHz.
Moreover, the detection of different species of nuclei (other than just $^{13}\mathrm{C}$ and $^1\mathrm{H}$) was presented in a (now) standard spectrometry measurement  \cite{DeVience2015} and more remarkably in scanning probe microscopy of a Teflon sample above a stationary NV center \cite{HaeberleT.2015}. These two last works, together with another scanning probe proof-of-principle demonstration of nuclear spin scanning probe microscopy \cite{RugarD.2014} have finally demonstrated the true power of the NV center in diamond as a local sensor for NMR.
\begin{figure*}[ht]
	\centering
	\includegraphics[width = 0.75\textwidth]{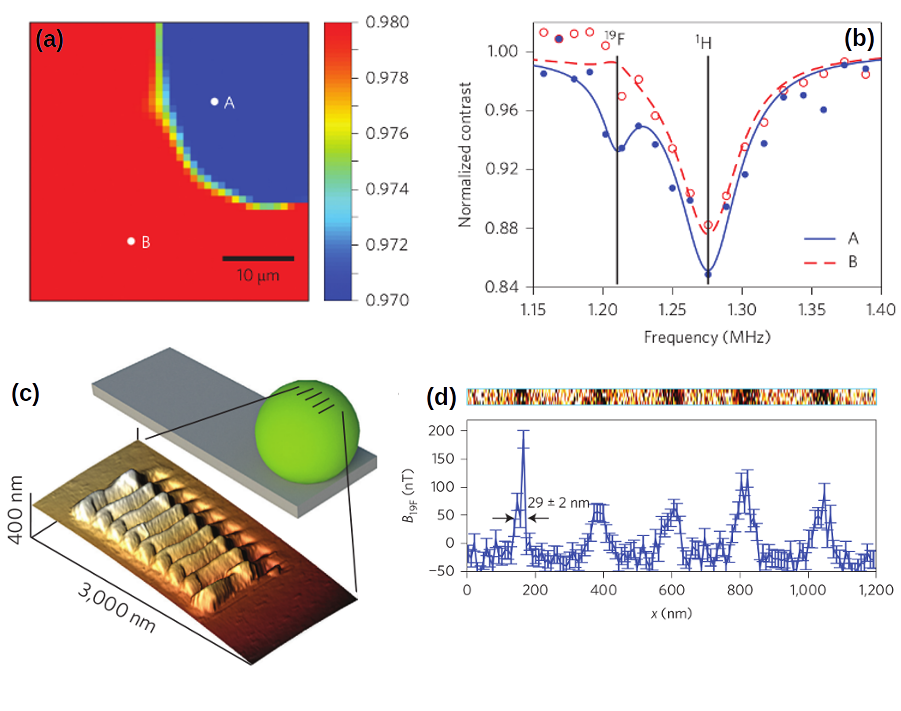}
	\caption{NMR imaging with the NV center in diamond: (a) Signal from $^{19}\mathrm{F}$ nuclear spin density in the fluorinated sample
(PFOS/POSF) within $\sim 20$ nm of the surface. Color indicates NMR contrast, with blue representing a deep $^{19}\mathrm{F}$ NMR contrast dip and hence high fluorine concentration while red represents no $^{19}\mathrm{F}$ NMR signal. This measurement was taken by an ensemble of NVs, see Ref.~\citealp{DeVience2015}. (b) NV NMR spectra for two points (A and B) of the image in (a), fit with Lorentzian curves. (c) A calibration grating is engraved into a sample of $^{19}\mathrm{F}$-rich Teflon AF (green sphere). (d) $^{19}\mathrm{F}$ NMR image, taken in a single-NV scanning probe technique, Ref.~\citealp{HaeberleT.2015}.}
	\label{figure9}
\end{figure*}
In Fig.~\ref{figure9}, two completely different methods of using defects in diamonds to provide magnetic resonance imaging are displayed. Figs.~\ref{figure9}a and b show MRI imaging using an ensemble of NVs in a widefield magnetometer setup, while Figs.~\ref{figure9}c and d provide the first experimental proof-of-principle of a scanning, nanoscale, NV-based MRI probe. In both cases the nucleus used was fluorine.

\section{Prospects and outlook}\label{summary}
With its point-like size, proximity and operation at room temperature, the current state-of-the-art for the NV center in diamond has now finally reached the stage where it is possible to detect single electron spins from an external target sample. The detection of single nuclear spins \cite{Mueller2014, PhysRevLett.113.197601} seems to be on the brink of accomplishment. With those two seminal milestones being achieved, this certainly enables the application of single-spin ESR and NMR detection to systems ranging from molecules to high-T$_\mathrm{C}$ superconductors. There is a constant drive to further push the limits of sensitivity and also the spectral resolution. Here we discuss several of those efforts.
\subsection{High resolution NMR and QIP}
Since the NV center in diamond is a true single two-level system (under a finite magnetic field which separates the degeneracy between the $\ket{m_s = 1}$ and the $\ket{m_s=-1}$ states), all quantum-information-processing (QIP) protocols can be applied to it in an almost straight-forward way. For a two-level system\footnote{This is a general result for a simple, ideal two-level system evolving under $H = \hbar\left(  \begin{array}{cc} 0 & \tfrac{1}{2}\Omega(t) \\ \tfrac{1}{2}\Omega(t) & \Delta-\tfrac{1}{2}i\Gamma \end{array}\right)$, where $\Delta$ is the detuning from resonance}, the limit to spectral linewidth, $\Delta\nu$, lies with two parameters: The driving power, which in our case is the Rabi frequency, $\Omega$, and the typical relaxation time of the system, or $1/\Gamma = T_1$, such that
\begin{equation*}
	\frac{\Delta\nu}{\nu} \propto \frac{1}{\Omega\cdot T_1}
\end{equation*}
This immediately warrants the use of different schemes to increase both the coherence time, by adapting, for example, quantum error correction protocols \cite{Unden2016}, quantum Fourier transform and quantum memory for storing the state of the electron spin (\cite{Laraoui2013, Zaiser2016, Pfender2016} and also to go beyond the limitation of the NV's $T_1$ relaxation time and achieve, as a result, record spectral resolution.
\subsection{Chemical shifts}
One of applicative uses of NMR is to detect the chemical shifts from the Larmor frequency due to intra-molecular bonds and structure. These shifts are measured in ppm - for the same type of nucleus, the shift can range from a few to several tens of ppm.
Two very recent works \cite{Pfender2016, Rosskopf2016} have reported a linewidth of 16 Hz and 39 Hz, respectably. The former was attained at a magnetic field of 1.5 Tesla for $^{13}$C nuclei ($\omega_L = 16.04$ MHz), which is effectively a resolution of less than 2 ppm. Apart from the intrinsic $^{13}\mathrm{C}$ nuclei in the diamond lattice, an oil containing perfluoropolyether (PFPE) molecules was cast on the surface of the diamond, and the $^{19}\mathrm{F}$ shift was determined to be 8 kHz $\pm$ 3.9 kHz, in line with the literature value of 10 kHz. Both approaches use the NV's nitrogen nucleus as quantum memory for storing the amplitude and phase of the NV's electron spin, which enable a longer free evolution time for sensing nuclei in the target sample. Their advantage over the lock-in scheme \cite{Staudacher2013} or the active-RF scheme \cite{Mamin2013} is that the NV is not in a superposition of $\ket{0}$ and $\ket{1}$ spin states, but rather at the $\ket{0}$ state, such that it is not sensitive to hyperfine interactions with nearby nuclei. In addition, the RF pulses which are used in that scheme are performed, once again, when the NV is in its $\ket{0}$ state. Such linewidths already approach or are on the same level as the capabilities of commercial NMR machines, with the distinct feature, of course, of detecting a very small number of nuclear spins.
\subsection{The search for equivalent defects - SiC}
As stated above, The NV center in diamond is not the only defect which is useful for magnetic resonance detection. The five criteria, as they were well devised in Ref.~\citealp{Weber2010}, pose strict requirements on possible candidates: (1) a paramagnetic and long-lived bound state with an energy splitting between two sublevels of its spin states; (2) an optical pump cycle that polarizes the system; (3) spin-state dependent fluorescence; (4) optical transitions that do not interfere with the host material's electronic states; and (5) the energy difference between bound states should be larger than the thermal excitation. There are also some ideal requirements for the host material, chief among them is the wide band-gap. One of the most promising candidates, which fulfils these criteria is the silicon-vacancy in silicon-carbide (4H-SiC) \cite{Koehl2011, Widmann2014a}. are showing promising properties for sensing applications.
\subsection{Single nuclear spin detection}\label{singlespin}
For the unambiguous detection of an individual nuclear spin, it is necessary to enter the regime where the interaction of said spin with the sensor (in our case the NV center in diamond) is stronger than its interaction with the surrounding spin bath. This can be defined as strong coupling \cite{Mueller2014}. A clear illustration of the different coupling regimes has already been shown in Figs.~\ref{figure7}e-h.
As opposed to the weak-coupling regime, where the signal is in fact magnetic noise from nearby nuclei, in the strong-coupling regime, we would be measuring a signal proportional to the number of detected spins. Moreover, when detecting an external nuclear spin with the NV's electron spin, it is possible to verify whether it is a single spin or not by looking for a coherent joint rotation due to entanglement between the two \cite{Taminiau2012, Zhao2012}. The combination of working in the strong-coupling regime by, for example, diluting the spin sample to such an extent that the interaction with the NV is stronger than that with neighboring nuclei, with coherent joint rotation type measurement gives, then, the condition and method to use single defects in diamond to detect and coherently interact with single nuclear spins.
\section*{Acknowledgements}
We would like to thank Matthias Pfender, Nabeel Aslam, Sebastian Zaiser, Thomas H\"{a}berle and Dominik Schmid-Lorch for critical and fruitful discussions and Durga Dasari Bhaktavatsala Rao for proof-reading the manuscript. This work was partially funded by DARPA (QuASAR) and the EU (DIADEMS). A.\ F.\ acknowledges financial support from the Alexander von Humboldt Foundation.
\section*{References}

\bibliography{NVSPM}

\end{document}